\documentclass[preprint,numbers,sort&compress,3p,twocolumn]{elsarticle}

\usepackage{amssymb}
\usepackage{xspace}

\usepackage{lineno}

\newcommand{\un}[1]{\ensuremath{\, \mathrm{#1}}}

\newlength{\delimw}
\newlength{\rxfc} 

\newcounter{temp}

\newcommand{\icecube}{\textsc{IceCube}\xspace}
\newcommand{\inice}{\textsc{InIce}\xspace}
\newcommand{\icetop}{\textsc{IceTop}\xspace}

\newcommand{\kascade}{\textsc{Kascade}\xspace}

\newcommand{\auger}{\textsc{Auger}\xspace}

\newcommand{\casamia}{\textsc{Casa-Mia}\xspace}

\newcommand{\magic}{\textsc{Magic}\xspace}
\newcommand{\hess}{\textsc{H.E.S.S.}\xspace}

\newlength{\figwidth}
\newlength{\figsep}
\newlength{\pba} \newlength{\wa} 
\newlength{\pbb} \newlength{\wb} 
\newlength{\pbc} \newlength{\wc} 
\newlength{\ha} 

\newcommand{\onefig}[3][tbp]{
  \begin{figure}[#1]
    {\center
      \includegraphics[clip=true, width=\figwidth]{#2}\caption{#3}
    }
  \end{figure}
}

\newlength{\thinlinewidth} 
\newlength{\thicklinewidth} 

\usepackage[compact]{titlesec}

\begin{document}

\begin{frontmatter}



\title{A Radio Air-Shower Test Array (RASTA) for IceCube}


\author[ubonn]{Sebastian B\"oser\corref{cor1}}
\cortext[cor1]{corresponding author\newline\hspace*{1.7\parindent}{\it Email address: }{\tt sboeser@physik.uni-bonn.de}}
\author{for the IceCube collaboration}

\address{Physikalisches Institut, Universit\"at Bonn, 53113 Bonn}

\begin{abstract}
In this paper we explore the possibility to complement the cosmic ray physics
program of the \icecube observatory with an extended surface array of radio
antennas. The combination of air-shower sampling on the surface and muon
calorimetry underground offers significant scientific potential: the neutrino
sensitivity above the horizon can be enhanced by vetoing air-showers on the
ground, photon-induced air-showers can be identified by their small muon
component and the coincident measurement of the particle density on the surface
and the muon component gives useful information on the composition of the
primary flux.

All of these analyses are pursued with the existing \icetop array. However, the
\icetop footprint is small compared to the acceptance of the \inice sensor array,
which severely limits the solid angle for coincident measurements, calling for
an extended surface air-shower detector. As demonstrated by the LOPES
experiment, measuring air-showers through their geosynchrotron emission has
become a viable and cost-efficient method. The science case for the RASTA
project - a dedicated radio array seeking to exploit this method at the South
Pole - is presented.  \end{abstract}


\end{frontmatter}


\section{Motivation}
After first being discovered in the late 1960s \cite{AllanNature}, radio
emission of air showers has again received increasing interest in the past
years. This is mainly due to the achievements in information technology,
which open the possibility to use phased arrays of radio antennas in
combination with digital beam-forming. A first proof of this technology has been
provided by the {\sc Lopes} experiment \cite{Falcke:2002tp,Falcke:2005tc}. Its main
advantage are low costs and simple design of the radio antennas as
compared to classical air shower detector elements (such as scintillators or
photomultipliers) and the very large field of view provided by the phased array
(usually close to $2\pi$). This technology is now not only being developed as an
extension at many existing air shower experiments ({\sc Kascade, Auger,
Tunka,...}), but also forms the key building block of
{\sc Lofar} -- a multi-site, multi-purpose radio detector spanning half of
Europe \cite{Lofar:2005wt}.\\
In this paper, we will first outline the science case for an extension of the
\icecube observatory with an extended $\O(10\un{km^2})$ radio array, then describe
the current status and finally outline a roadmap towards a full detector.

\section{Science Case}
The \icecube observatory \cite{Berghaus:2008bk} at the South Pole was originally designed and intended
for detection
of $\un{TeV}-\un{PeV}$ neutrinos penetrating Earth, while at the same time using
it as a shield against the abundant atmospheric muon flux. Beyond the original
anticipation, \icecube has by now become a $4\pi$ observatory. In
particular, the \icetop air-shower array \cite{Stanev:2009ce} on the surface,
expands the physics reach of \icecube in a threefold manner, as detailed in the
sections below.
\onefig{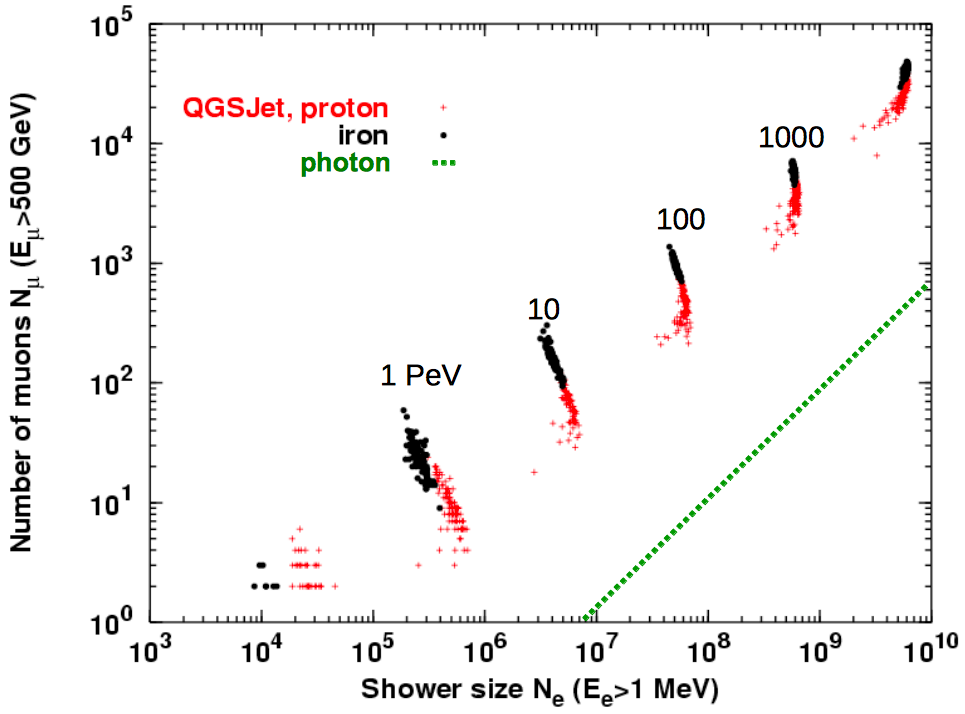}
{Detectable muons in \icecube vs. electrons in \icetop for different
energies and primaries. The dashed line shows an analytic calculation from
\citet{Halzen:1996ik}.\label{fig:compo}}
\subsection{Composition\label{sec:composition}}
Two different observables are currently employed to determine primary composition
of cosmic rays:
\begin{itemize}
\item the depth of the {\bf shower maximum $X_{max}$}, where on
average protons of the same energy will penetrate deeper in the atmosphere and
show larger variations than heavier nuclei. This is used to measure composition
e.g. with the fluorescence detectors in \auger.  \item the {\bf electron over
muon ratio}, where heavier primaries will
cause higher muon multiplicities (c.f. fig.\ref{fig:compo}), as exploited e.g. by
\kascade.
\end{itemize}
Significant statistics of coincident \inice and \icetop events have been
collected and are being analyzed. Due to the strong correlation of the muon and
electron flux with the total primary energy and inclination of the shower,
very good control of systematic effects in the reconstruction of the events is
required \cite{Klepser:2008pm}.

For air showers that are coincidentally detected by an additional future large
radio array, both -- the muon and the electron component of the shower -- will
be measured by more than one type of detector, allowing an overconstrained
determination of both components. Furthermore, simulations show that the steepness
of the lateral distribution of the geosynchrotron signal will depend on the
depth of the shower maximum \cite{Huege:2005rd,deVries:2010ti} -- an effect that will be
enhanced for higher observation levels \cite{Huege:2006kd}. Radio detection may
thus provide a view of the shower development at high altitudes. It should be noted,
though, that the simulation of the geosynchrotron emission is still a developing
field. The reconstruction of the radio signals will thus have to be validated
against the \icetop measurements first.

The benefit will go beyond the independent systematic
errors of the three detector types on the shower energy and directional
reconstruction. With the increased statistics provided by a large array, the
energy range that is covered with good statistics will be extended to the
$\un{EeV}$ regime, overlapping and complementing not only the measurements
provided by the \kascade but also of the \auger experiments. Using the
additional information on the shower maximum, a combined measurement will
eventually maximally exploit all handles on primary particle composition that
are currently available to earth-bound detectors. \\

\subsection{Neutrinos}
At energies up to $\O(\un{PeV})$, the neutrino field of view of \icecube is
limited to the $2\pi\un{sr}$ of the northern hemisphere due to the abundant flux
of atmospheric muons from air showers from above. This flux shows a much steeper
spectrum than the expected signal flux. At energies above several
$\O(100)\un{TeV}$ the neutrinos themselves get absorbed in Earth, so that
the search must be extended to the southern hemisphere.  Still, even at these
higher energies, the dominant background is from unresolved
muon bundles induced by air showers, which can not be easily distinguished from
single muons in \icecube. 
While \icetop in principle offers the possibility to detect the air
showers from which these muons stem, its vetoing power is limited by its small
size compared to the aperture available to \icecube to around $0.25\pi\un{sr}$. A radio array with
sufficient lateral extent can significantly increase the field of view in which
air shower induced muons can be vetoed. An estimate shows that the
sensitivity to the GZK neutrino flux may be increased by a factor of more than
three for an array extending to $3\un{km}$ around the center of \icetop
\cite{Jan:2010}.

\subsection{Ultra-high energy gamma rays}
Photon induced air showers will show a muon flux that is a factor of $\approx 100$
lower than for hadron induced showers \cite{Drees:1988wu} (c.f. fig.\ref{fig:compo}). At the same time, the flux of
muons induced by hadron showers is energetic enough that \icecube will have
a nearly $100\%$ detection efficiency for showers of energies above
$10^{15}\un{eV}$ if the shower axis intersects the geometric volume. This can
be exploited to discriminate hadronic air showers against electromagnetic
showers, turning the observatory in a $\un{km^2}$-scale $\un{PeV}$ photon
detector. 

While at $\un{PeV}$ energies, the UHE gamma horizon is limited to our own galaxy
due to absorption on the cosmic microwave background, both the \hess and \magic
experiment report unbroken $E^{-2}$ spectra for some sources within our galaxy
\cite{Abdo:2007ad,Hess:2007}.  Photons have been detected from the same sources with energies
up to energies of $90\un{TeV}$ \cite{Abdo:2007ad}. If point sources of gamma
rays with unbroken spectra up to $\un{PeV}$ exist, they should generate $\O(10)
\un{evt}\un{yr^{-1}}$ in an array as described above.  Non-excluding limits on
the diffuse photon fraction in cosmic rays have been set previously at
$<10^{-4}$ for the energy range up $5\cdot10^{16}\un{eV}$ by the \casamia
collaboration\cite{Chantell:1997gs} and at $<10^{-1.6}$ for energies above
$2\cdot10^{18}\un{eV}$ by \auger\cite{Auger:2009qb}.  Ongoing
analyses \cite{BuitinkPriv} with \icetop show that a comparably high sensitivity can be
reached, but only for the limited section of the galactic plane covered by the
\icetop-\inice field-of-view. 

Again a future radio array will enhance the physics reach for UHE gamma
detection. As the extent of a future radio array will be much larger than the
extent of \icetop, sources at higher declination -- including the entire
galactic plane -- can be probed. At the same time, the larger collection area
will increase event statistics, and will thus bridge the gap in energy
between \casamia and \auger. Finally, as is discussed in section
\ref{sec:composition}, the
radio signal can additionally provide a handle on the depth of the shower
maximum, hence further improve the photon identification in this search.

\onefig[t]{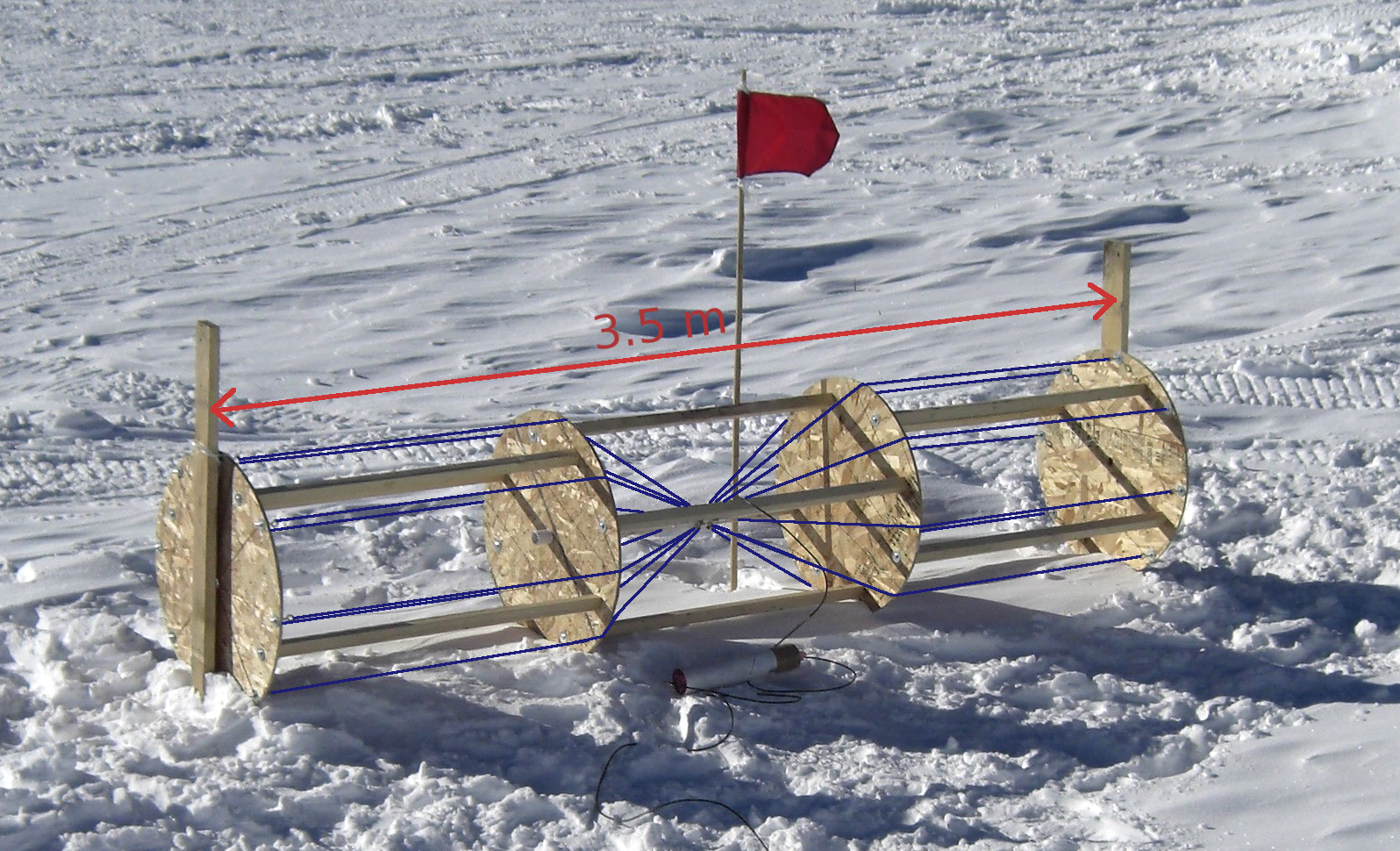}
  {Custom {\em Fat Wire-Dipole} (FWD) antenna\label{fig:antenna}}
\section{Current status and experimental setup}
A set of dedicated measurements has been performed to explore the possibility of
such a radio array in Antarctic conditions~\cite{Jan:2010}. A {\em fat wire-dipole} antenna
depicted in Figure \ref{fig:antenna} has been developed with the aim of a high
and uniform broadband response in the $\un{MHz}$-regime while limiting antenna
dimensions and cost.

Two of these antennas were deployed at South Pole in the polar season 2008/2009
around the SPASE building and were read out using two channels of the
pre-existing RICE DAQ \cite{Kravchenko:2007fy}. The sensitivity of the deployed instrumentation was severely
limited by cable losses as only two $600\un{m}$ and $500\un{m}$ pieces of legacy
$75\un{\Omega}$ coaxial cable were available for use. In the frequency range
between 50 and 100 MHz signal transmission losses vary from 20-30 dB.
In absence of a dedicated trigger for the surface antennas the $> 2
\un{\mu{}s}$ cable delays also resulted in signals that were often truncated by the
limited buffer length of the RICE waveform capture.
Nevertheless, the essential pre-conditions for establishing an enlarged
footprint at South Pole have been achieved via two fundamental performance
milestones.

\onefig[t]{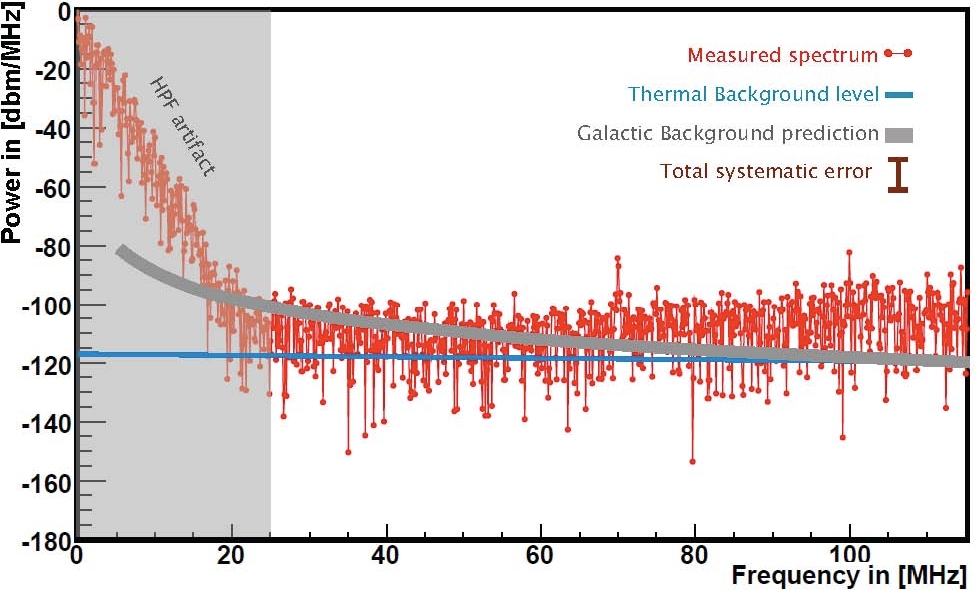}
  {$8\un{\mu{}s}$-sweep noise spectrum from fat wire-dipole\label{fig:noisespectrum}}
1) Analysis of the spectrum shows the noise floor being limited by galactic
and thermal contributions. Figure \ref{fig:noisespectrum} shows the background spectrum
from a single $8\un{\mu{}s}$ sweep corrected for the attenuation effects of
cable, filter, and amplifier. The strong increase below
$25\un{MHz}$ is an artefact of the high-pass filter being used. As in a previous short-term
measurement during the 2007/2008 season, the spectrum lacks the typical
monochromatic features from TV stations and other anthropogenic sources observed
in other more inhabitated areas. 

2) To test the ability of surface antennas to identify sources at South Pole, a set of
calibration pulser runs were taken in January 2009 using a RICE transmitter
located at the $500\un{m}$ distant MAPO
building. Taking advantage of the nanosecond time resolution of the RICE DAQ,
the source position has been reconstructed using only information from the two
fat wire-dipoles plus two auxilliary dipole antennas at
the roof of MAPO building. The vast majority of the events reconstruct to within
$\sim 2\un{m}$ of the actual pulse location.

\section{Roadmap}
Here we outline a roadmap to assess the viability of an extended air-shower
array for geosynchrotron radiation as described above. Considering the
restricted hardware deployment cycle due to the remote location, a
staged proposal spanning three years has been made.
The setup in the first year will be aimed towards unambiguously establishing the
detection of air-showers above the background radio noise and thus provide a
proof-of-viability for this detection technique in the Antarctic
environment. Using a conservative estimate based on REAS2 simulations
\cite{Jan:2010}, air-showers with a primary energy of $E_{prim} > 10^{17}
\un{eV}$ should be detectable at $> 5\sigma$ above the ambient noise level at a
distance of $125\un{m}$.
Antennas will be deployed in pairs in orthogonal
polarization. Four such pairs at a baseline of $55\un{m}$ and requiring a four-fold coincidence
will provide an effective area of $3\cdot10^{4}\un{m^2}$. Using the charged cosmic ray flux as measured by the
Kascade experiment \cite{Ulrich:2002}, each of these clusters will detect $\sim 8
\un{events/day}$. Even allowing for considerable inefficiencies (e.g. due to
anthropogenic backgrounds), some $10^{3}$ events will be accumulated per year.

While this primary setup will heavily rely on
commercially available components and existing infrastructure, the setup in the second season
should employ all of the key technologies that are required for a
several-$\un{km}^2$ array, hence will also allow to demonstrate the scalability
of the approach. In particular these include each antenna pair to be read out by
an independent, custom DAQ board deployed in close proximity. Connecting the
stations in a peer-to-peer grid, in which each of the boards generates threshold
triggers and communicates them to neighboring antennas will allow to establish
local coincidences. Local storage of the waveform information with a buffer
depth of several $100\un{ms}$ will be required to read out the array following a
global trigger either generated internally from the radio
array or externally from \icecube or \icetop. With a significantly larger number
of antennas on an enlarged footprint, this second stage will collect a
data sample large enough to allow detailed verification of the array
performance and shower reconstruction. 

The third year will be dedicated to investigation of long-term systematics. With
data accumulated over two years, potential environmental changes (such as
compactification of the snow around the antennas) or degradation of the antenna
system will be studied in detail. Any needed in-situ recalibration of the
antennas and comparison to previous calibrations would be conducted in this
campaign. For this purpose a radio-frequency transmitter can be deployed at
different locations in the array.

\section{Summary}
\icecube{}'s core mission includes several cosmic ray initiatives. While combined analysis of
\icetop and \inice data is being used to target cosmic-ray composition, there will continue to be
uncertainties driven by the ability to completely characterize individual events.
An extended air-shower radio array using geosynchrotron radiation will significantly enhance the
acceptance of \icecube for the veto and gamma-ray missions. It will also provide complementary
information for shower reconstruction. The combined analysis of
\inice{}/\icetop{}/RASTA events
will permit cross-calibration of all detection techniques and improved composition measurements.
The proposed three year development plan will allow to establish and test all
the key ingredients required towards reaching these goals.





\bibliographystyle{model1a-num-names}
\bibliography{radio}

\begin{thebibliography}{20}
\expandafter\ifx\csname natexlab\endcsname\relax\def\natexlab#1{#1}\fi
\providecommand{\bibinfo}[2]{#2}
\ifx\xfnm\relax \def\xfnm[#1]{\unskip,\space#1}\fi
\bibitem[{Allan et~al.(1969)Allan, Clay, Jones et~al.}]{AllanNature}
\bibinfo{author}{H.~R. Allan}, \bibinfo{author}{R.~Clay},
  \bibinfo{author}{A.~Jones}, et~al., \bibinfo{journal}{Nature}
  \bibinfo{volume}{222} (\bibinfo{year}{1969}) \bibinfo{pages}{635--637}.
\bibitem[{Falcke and Gorham(2003)}]{Falcke:2002tp}
\bibinfo{author}{H.~Falcke}, \bibinfo{author}{P.~Gorham},
  \bibinfo{journal}{Astropart. Phys.} \bibinfo{volume}{19}
  (\bibinfo{year}{2003}) \bibinfo{pages}{477--494}.
\bibitem[{Falcke et~al.(2005)}]{Falcke:2005tc}
\bibinfo{author}{H.~Falcke}, et~al., \bibinfo{journal}{Nature}
  \bibinfo{volume}{435} (\bibinfo{year}{2005}) \bibinfo{pages}{313--316}.
\bibitem[{Br\"uggen et~al.(2005)}]{Lofar:2005wt}
\bibinfo{author}{M.~Br\"uggen}, et~al., \bibinfo{title}{German {LOFAR}; white
  paper}, \bibinfo{howpublished}{Max-Planck-Institut f\"ur Radioastronomie},
  \bibinfo{year}{2005}.
\bibitem[{Berghaus(2009)}]{Berghaus:2008bk}
\bibinfo{author}{P.~Berghaus}, \bibinfo{journal}{Nucl. Phys. Proc. Suppl.}
  \bibinfo{volume}{190} (\bibinfo{year}{2009}) \bibinfo{pages}{127--132}.
\bibitem[{Stanev and {for the IceCube Collaboration}(2009)}]{Stanev:2009ce}
\bibinfo{author}{T.~Stanev}, \bibinfo{author}{{for the IceCube Collaboration}},
  \bibinfo{journal}{Nucl. Phys. Proc. Suppl.} \bibinfo{volume}{196}
  (\bibinfo{year}{2009}) \bibinfo{pages}{159--164}.
\bibitem[{Halzen et~al.(1997)Halzen, Stanev, and Yodh}]{Halzen:1996ik}
\bibinfo{author}{F.~Halzen}, \bibinfo{author}{T.~Stanev},
  \bibinfo{author}{G.~B. Yodh}, \bibinfo{journal}{Phys. Rev.}
  \bibinfo{volume}{D55} (\bibinfo{year}{1997}) \bibinfo{pages}{4475--4479}.
\bibitem[{Klepser and {for the IceCube Collaboration}(2008)}]{Klepser:2008pm}
\bibinfo{author}{S.~Klepser}, \bibinfo{author}{{for the IceCube
  Collaboration}}, \bibinfo{journal}{Proc. of ECRS}  (\bibinfo{year}{2008}).
\bibitem[{Huege and Falcke(2005)}]{Huege:2005rd}
\bibinfo{author}{T.~Huege}, \bibinfo{author}{H.~Falcke},
  \bibinfo{journal}{Astropart. Phys.} \bibinfo{volume}{24}
  (\bibinfo{year}{2005}) \bibinfo{pages}{116--136}.
\bibitem[{de~Vries et~al.(2010)}]{deVries:2010ti}
\bibinfo{author}{K.~D. de~Vries}, et~al., \bibinfo{journal}{arXiv:1008.3308v1}
  (\bibinfo{year}{2010}).
\bibitem[{Huege et~al.(2007)Huege, Ulrich, and Engel}]{Huege:2006kd}
\bibinfo{author}{T.~Huege}, \bibinfo{author}{R.~Ulrich},
  \bibinfo{author}{R.~Engel}, \bibinfo{journal}{Astropart. Phys.}
  \bibinfo{volume}{27} (\bibinfo{year}{2007}) \bibinfo{pages}{392--405}.
\bibitem[{Auffenberg(2010)}]{Jan:2010}
\bibinfo{author}{J.~Auffenberg}, Ph.D. thesis, Univ. of Wuppertal,
  \bibinfo{year}{2010}.
\bibitem[{Drees et~al.(1989)Drees, Halzen, and Hikasa}]{Drees:1988wu}
\bibinfo{author}{M.~Drees}, \bibinfo{author}{F.~Halzen},
  \bibinfo{author}{K.~Hikasa}, \bibinfo{journal}{Phys. Rev.}
  \bibinfo{volume}{D39} (\bibinfo{year}{1989}) \bibinfo{pages}{1310}.
\bibitem[{Abdo et~al.(2007)}]{Abdo:2007ad}
\bibinfo{author}{A.~A. Abdo}, et~al., \bibinfo{journal}{Astrophys. J.}
  \bibinfo{volume}{664} (\bibinfo{year}{2007}) \bibinfo{pages}{L91--L94}.
\bibitem[{Djannati-Atai et~al.(2007)}]{Hess:2007}
\bibinfo{author}{A.~Djannati-Atai}, et~al., \bibinfo{journal}{Proc. of 30th
  ICRC, Merida, Mexico}  (\bibinfo{year}{2007}).
\bibitem[{Chantell et~al.(1997)}]{Chantell:1997gs}
\bibinfo{author}{M.~C. Chantell}, et~al., \bibinfo{journal}{Phys. Rev. Lett.}
  \bibinfo{volume}{79} (\bibinfo{year}{1997}) \bibinfo{pages}{1805--1808}.
\bibitem[{Abraham et~al.(2009)}]{Auger:2009qb}
\bibinfo{author}{J.~Abraham}, et~al., \bibinfo{journal}{Astropart. Phys.}
  \bibinfo{volume}{31} (\bibinfo{year}{2009}) \bibinfo{pages}{399--406}.
\bibitem[{Buitink(2010)}]{BuitinkPriv}
\bibinfo{author}{S.~Buitink}, in: \bibinfo{booktitle}{talk at \icecube
  collaboration meeting}.
\bibitem[{Kravchenko et~al.(2007)}]{Kravchenko:2007fy}
\bibinfo{author}{I.~Kravchenko}, et~al., \bibinfo{journal}{arXiv:0705.4491}
  (\bibinfo{year}{2007}).
\bibitem[{Ulrich et~al.(2003)}]{Ulrich:2002}
\bibinfo{author}{H.~Ulrich}, et~al., \bibinfo{journal}{Nucl. Phys. B.
  Proceedings} \bibinfo{volume}{122} (\bibinfo{year}{2003})
  \bibinfo{pages}{218--221}.

\end{thebibliography}







\end{document}